\documentclass[12pt, letterpaper]{article}
\usepackage{jheppubmodnew}
\pdfoutput=1
\usepackage{hyperref}
\usepackage{psfrag}
\usepackage{array}
\usepackage{amssymb}
\usepackage{amsmath}
\usepackage{amsthm}
\usepackage{graphicx}
\usepackage{enumitem}
\usepackage{diagbox}
\usepackage[labelsep=quad]{subcaption}
\usepackage{cite}

\newcommand{\psine}{\Psi^{\text{ne}}} 

\def\psinesub[#1]{\psine_{#1}}

\newcommand{\cop}[1]{#1}
\newcommand{\al}{\cop{A}} 
\newcommand{\alset}{{\cal A}} 

\newcommand{\op}{{\cal O}} 

\newcommand{\sent}{S^{\text{ent}}}

\def\Oright[#1]{\op_{#1}} 
\def\Oleft[#1]{\op_{L#1}} 
\def\anc{\cop{a}}

\def\an[#1]{\anc_{#1}} 
\def\aleft[#1]{\cop{a}_{L#1}} 
\def\arelr[#1]{\cop{a}^{\text{rel}}_{R#1}} 

\def\alen[#1]{\al_{\text{L}, #1}} 

\def\enrange[#1,#2]{{\cal R}_{#1}}
\def\dimrange[#1,#2]{{\cal D}_{#1}}
\def\hilb[#1]{{\cal H}_{#1}}

\def\projrange[#1, #2]{\cop{P}_{#1}}
\def\projh[#1]{\cop{P}_{\hilb[#1]}}

\def\proj[#1]{\cop{P}_{#1}}

\def\tarelr[#1]{\widetilde{a}^{\text{rel}}_{R#1}}
\def\coeff[#1]{\alpha_{#1}}

\def\pb[#1,#2]{\{#1, #2\}}
\def\deb[#1,#2]{[#1,#2]_{\text{D.B.}}}

\def\tr{{\rm Tr}}

\def\Or[#1]{{\text{O}}\left({#1}\right)}
\def\dotl[#1,#2]{\left\langle #1,\, #2 \right\rangle}
\def\dotlb[#1,#2]{\left\langle #1,\, #2 \right\rangle}
\def\dotlm[#1,#2]{\left[ #1,\, #2 \right]}
\def\dotp[#1,#2]{(\vect{#1} \cdot\vect{#2})}
\def\aff[#1,#2]{\hat{#1}(#2)}

\def\n4sym{{\cal N}=4 SYM}
\def\>{\rangle}
\def\<{\langle}
\def\weight[#1,#2,#3]{\{(#1),#2,#3\}}
\def\ads[#1]{$\text{AdS}_{#1}$}

\def\rtors[#1]{r_{*#1}}

\newcommand{\be}{\begin{equation}}
\newcommand{\ee}{\end{equation}}
\newcommand{\ba}{\begin{align}}
\newcommand{\ea}{\end{align}}
\newcommand{\bs}{\begin{split}}

\def\sess\end{split}
\newcommand{\vect}[1]{{\vec{#1}}}

\def \scrip{{\cal I}^{+}}
\def \scrim{{\cal I}^{-}}

\def\alcut[#1]{{\cal A}_{#1, \epsilon}}
\def\alseg[#1,#2]{{\cal B}_{#1, #2}}
\def\alscripast{{\cal A}_{-\infty, \epsilon}}
\def\abdry{{\cal A}_{\text{bdry}}}

\def\supcharge[#1]{\{#1\}}

\def\projsupeig[#1]{{\cal P}_{{\ell, m}}[{#1}]}
\def\transop[#1, #2]{T_{\{#1\}, \{#2\}}}
\def\supket[#1]{|\{#1\} \rangle}
\def\supbra[#1]{\langle \{#1\} | }

\newcommand{\rcomp}{\overline{R}_{\epsilon}}

\title{Failure of the split property in gravity and the information paradox}
\author{Suvrat Raju}
\affiliation{International Centre for Theoretical Sciences, Tata Institute of Fundamental Research, Shivakote, Bengaluru 560089, India.}
\emailAdd{suvrat@icts.res.in}
\date{}
\abstract{
In an ordinary quantum field theory, the ``split property'' implies that the state of the system can be specified independently on a bounded subregion of a Cauchy slice and its complement.  This property does not hold for theories of gravity, where observables near the boundary of the Cauchy slice uniquely fix the state on the entire slice. The original formulation of the information paradox explicitly assumed the split property and we follow this assumption to isolate the precise error in Hawking's argument. A similar assumption also underpins the monogamy paradox of Mathur and AMPS. Finally the same assumption is used to support the common idea that the entanglement entropy of the region outside a black hole should follow a Page curve. It is for this reason that computations of the Page curve have been performed only in nonstandard theories of gravity, which include a nongravitational bath and massive gravitons. The fine-grained entropy at $\scrip$ does not obey a Page curve for an evaporating black hole in standard theories of gravity but we discuss possibilities for coarse graining that might lead to a Page curve in such cases.}

\begin{document}
\maketitle
It is sometimes possible to make progress simply by discarding an incorrect assumption. In this essay, we will discuss the incorrect assumption that the state of a system, in a standard theory of quantum gravity, can be specified independently on a bounded region and on its complement. 

It is not hard to see why this assumption, although incorrect, is appealing. First, such a property does hold for nongravitational quantum field theories, and can be formalized in terms of  the ``split property'' \cite{roos1970independence,buchholz1974product,Haag:1992hx} that we review below. The split property relies on the fact that, in nongravitational theories,  the algebra of observables on a Cauchy slice can be separated into a subalgebra associated with a bounded region and a commuting subalgebra associated with its complement.  Second, in classical gravity it is possible to find solutions that coincide outside a bounded region but differ inside \cite{Corvino:2003sp}. This has led to a common belief that, even in quantum gravity, observables outside a bounded region do not give us much information about the state inside the region except for the energy and a small number of conserved charges.

However, it has recently been shown \cite{Laddha:2020kvp,Chowdhury:2020hse,Chowdhury:2021nxw}, in several examples, that the split property fails in theories of quantum gravity. Instead such theories obey the principle of ``holography of information'', which states that all the information available on a Cauchy slice is also available near its boundary. In particular, this means that once all observables outside a bounded region have been specified there is {\em no freedom} to specify the state of the system inside the region. So the manner in which quantum gravity localizes information is the opposite of what the split property would suggest.

In this essay, we review how the split property has been assumed repeatedly in discussions of black-hole information --- sometimes tacitly and sometimes explicitly --- and this has often  led to paradoxes.  This assumption underpins Hawking's original argument for information loss \cite{Hawking:1974sw,Hawking:1976ra}. It  was also a key ingredient in the monogamy paradox formulated by Mathur \cite{Mathur:2009hf} and then elaborated by AMPS \cite{Almheiri:2012rt} that led to discussions of ``firewalls'' behind the horizon.  More interestingly, this assumption also underlies the idea that the entropy of Hawking radiation should obey a ``Page curve'' \cite{Page:1993df, Page:1993wv}. 

It is because this  assumption is false in standard theories of gravity that  computations of the Page curve \cite{Almheiri:2019psf,Penington:2019npb,Almheiri:2019hni,Almheiri:2019psy,Almheiri:2019yqk,Almheiri:2020cfm} have been carried out only in nonstandard theories of gravity that involve a nongravitational bath, contain a massive graviton and do not obey the Gauss law.  But these models of gravity do not provide an accurate understanding of the evaporation of black holes in standard gravity.  We will point out how simply discarding the assumption of the split property leads to a more robust resolution of Hawking's paradox.

This essay presents a brief summary of this perspective, with minimal technical details. We provide specific references to the literature in the text and refer the reader to the review  \cite{Raju:2020smc} for a more detailed discussion and additional references.  In section \ref{secsplitfailure}, we review the split property and its failure due to gravitational effects.   In section \ref{sechawkingerror}, we isolate the precise point at which the split property was assumed in Hawking's original argument for information loss and, in section \ref{secmonogamy} we describe the significance of this assumption in some  later refinements of the paradox.  We then explain, in section \ref{secpage}, how the same assumption led to the expectation that the entropy of the black hole exterior should follow the Page curve and review the significance of the nongravitational bath and massive gravitons in recent computations of the Page curve.   In section \ref{secpage}, we also discuss some possibilities for coarse-grained sets of observables that might lead to a Page curve in standard theories of gravity.  

However, one of the messages of this essay is that the failure of the split property in theories of gravity is not a bug that needs to be avoided. Rather, this unusual localization of quantum information in quantum gravity is an important physical lesson  that emerges from a study of black-hole information. By suitable manipulations, it is possible to construct systems that lead to a Page curve and make a theory of gravity resemble  nongravitational theories. But this exercise does not appear to be directly relevant for understanding the evaporation of black holes in a standard theory of gravity. In standard theories of gravity, black hole spacetimes store information in a manner that is more interesting than the Page curve would suggest.

\section{The split property and its failure in gravity \label{secsplitfailure}}

\subsection{Nongravitational quantum field theory \label{secsplitnongrav}} 
\begin{figure}
\begin{center}
\includegraphics[width=0.4\textwidth]{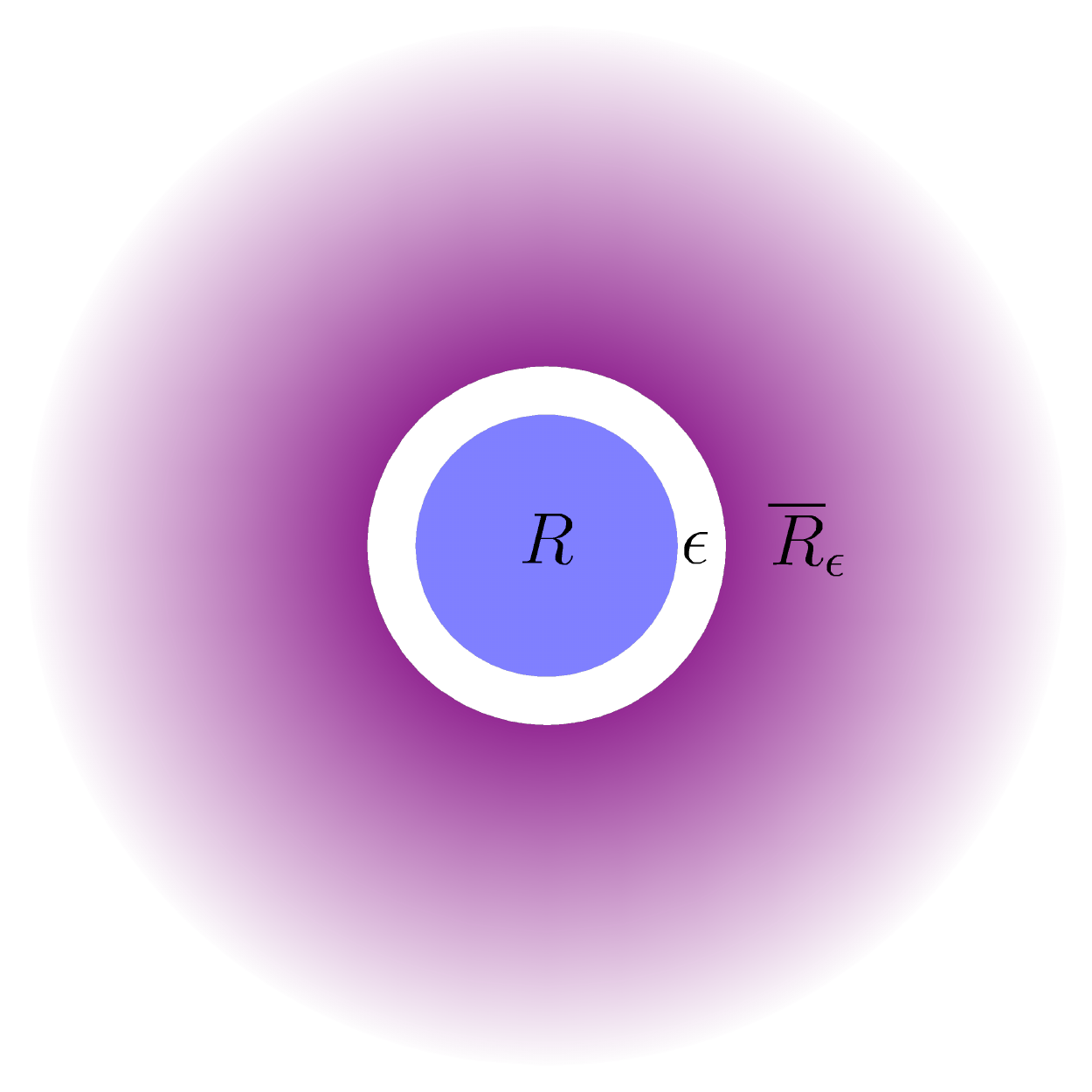}
\caption{\em The configuration of interest. We study a region $R$ (blue), surrounded by a collar region $\epsilon$ (white) and the complement of $R \cup \epsilon$, which is denoted by $\overline{R}_{\epsilon}$ (purple). The region, $\overline{R}_{\epsilon}$ extends to infinity. In a nongravitational theory, the state on $\overline{R}_{\epsilon}$ and $R$ can be specified entirely independently but in a theory of gravity the state on $\overline{R}_{\epsilon}$ completely fixes the state on $R$. \label{figbounded}}
\end{center}
\end{figure}

We start by briefly reviewing the split property in nongravitational quantum field theories \cite{roos1970independence,buchholz1974product,Haag:1992hx}. (See also the review \cite{Fewster:2016mzz}.) Consider a Cauchy slice in a noncompact spacetime. Let $R$ be a bounded region that is surrounded by a small ``collar region'', $\epsilon$. The complement of $R \cup \epsilon$ is denoted by $\rcomp$ as shown in Figure \ref{figbounded}.
  In a nongravitational theory, there is no obstacle to studying local operators, $\phi(x)$ that probe  the physics in the region $R$ and are labelled by a point $x \in R$.   As usual, it is possible to consider arbitrary polynomials of such operators  to generate the algebra of the region $R$ that we denote by $\alset$.
\be
\label{alsetgen}
\alset = \text{span~of}\{\phi(x_1), \phi(x_1) \phi(x_2), \ldots ,\phi(x_1) \phi(x_2) \ldots \phi(x_n), \ldots \}, \quad x_i \in R
\ee
Similarly, one may generate the algebra of observables in the region  $\overline{R}_{\epsilon}$ that we denote by $\overline{\alset}$. Then, in a nongravitational quantum field theory, microcausality implies that elements of these two algebras must commute pairwise
\be
[A, \bar{A}] = 0, \quad \forall A \in \alset, \bar{A} \in \overline{\alset}.
\ee

In the context above, the split property  states that if $\rho_1$ and $\rho_2$ are any two mixed states in the theory, which we represent as density matrices, then it is possible to find a ``split state'', $\rho$, that appears to be like $\rho_1$ in $R$ and like $\rho_2$ in $\overline{R}_\epsilon$. 
\be
\label{splitproperty}
\tr(\rho A \bar{A}) = \tr(\rho_1 A) \tr(\rho_2 \bar{A}), \qquad \forall A \in \alset, \bar{A} \in \overline{\alset}.
\ee
The split property tells us that, in a nongravitational theory, even if we are given the value of all observables in $\overline{R}_{\epsilon}$ we are completely ignorant about the state in $R$. 

This property just formalizes the notion that, in a lattice regularization, the Hilbert space of a nongravitational theory factorizes into a part associated with $R$ and a part associated with its complement. If the Hilbert space factorizes, we do not even need to introduce a collar region. Say the full Hilbert space can be written as a tensor product of a factor associated with $R$ and another factor associated with its complement $\overline{R}$: ${\cal H} = {\cal H}_R \otimes {\cal H}_{\overline{R}}$. A split state can then be generated simply by taking
\be
\label{factorizeddens}
\rho= \tr_{{\cal H}_{\overline{R}}}(\rho_1) \otimes \tr_{{\cal H}_{R}}(\rho_2); \qquad \text{split~state~in~a~factorized Hilbert space.}
\ee

In the continuum theory, if one does not include the collar region then a product state of the form above would have infinite energy. The collar region ensures that it is possible to preserve the very-short distance correlations across the boundary of $R$ and $\overline{R}_{\epsilon}$,  allow a smooth ``transition'' from the state $\rho_1$ in $R$ to the state $\rho_2$ in $\overline{R}_{\epsilon}$ and also ensure no correlations between $R$ and $\overline{R}_{\epsilon}$.  

If the reader finds the phrase ``split property'' unduly formal, the reader is free to think in terms of the more familiar but slightly imprecise idea that the Hilbert space ``factorizes'' into a part associated with a region and another part associated with its complement. Indeed, we will use this formulation below to make contact with the existing literature. 

The ``split property'', as written in the form \eqref{splitproperty}, also holds in nongravitational gauge theories. The existence of conserved charges does not modify \eqref{splitproperty}; measuring the charge in $\rcomp$ does not tell us about the charge contained in $R$ since an arbitrary amount of compensating charge can be stuffed into the collar region.

\subsection{Quantum gravity}

\subsubsection{Holography of information}
The split property fails in quantum gravity and is replaced by the principle of holography of information that we now review. We do not provide proofs of the results quoted below but include references to the relevant literature.

In a theory of gravity, it is simplest to consider the algebra of asymptotic observables. This is because the gauge redundancy of the theory only comprises diffeomorphisms that die off asymptotically. So asymptotic observables are gauge invariant.

First consider four-dimensional asymptotically flat space. Let us denote the algebra of operators at $\scrip$ by $\alset(\scrip)$.  (Similar statements hold for the algebra of operators on $\scrim$  and to avoid repetition, we only discuss $\scrip$.) The algebra $\alset(\scrip)$ is generated by the fields that describe the asymptotic fluctuations of the metric and of other dynamical fields. The fluctuations of the metric can themselves be divided into the Bondi news tensor $N_{AB}(u, \Omega)$ ($A,B$ run along the directions on $\scrip$)  and the Bondi mass aspect  $m(u, \Omega)$  both of which are labeled by a retarded time, $u$, and a point on the celestial sphere $\Omega$.\footnote{In \cite{Bousso:2017xyo}, it was argued that the mass aspect is not a well defined operator at finite retarded time. For what follows, we only need the mass aspect near $u \rightarrow -\infty$ which can be defined as described in Appendix B of \cite{Laddha:2020kvp}.} The algebra $\alset(\scrip)$ is then constructed by taking all possible polynomials in these operators precisely as in equation \eqref{alsetgen}.

Now, consider the subalgebra formed by operators near the past boundary of $\scrip$, which we denote by $\abdry$. (See Figure \ref{figbhevap}.) This is generated just as above except that we restrict to operators whose retarded time is in the range $(-\infty, -{1 \over \epsilon})$ for some choice of $\epsilon$.    Naively, one might have thought that $\abdry$ is a small subalgebra of $\alset(\scrip)$.  But it can be shown  any state of massless particles is {\em uniquely} specified by observables from this algebra. 
\be
\label{flatholinfo}
\tr(\rho_1 \al) = \tr(\rho_2 \al) ~ \forall \al \in \abdry \Leftrightarrow \rho_1 = \rho_2,
\ee
where $\rho_1$ and $\rho_2$ are any two mixed states in the massless Hilbert space.  The surprising aspect of the statement above is that equality of the expectation values is {\em not} imposed for all elements of $\alset(\scrip)$ but only for the elements of $\abdry$.  This result is proved as ``Result 1'' in \cite{Laddha:2020kvp} and we note that  $\abdry$ here is denoted by $\alscripast$ there.

We now turn to global AdS$_{d+1}$ whose conformal boundary is of the form $S^{d-1} \times R$ where the time coordinate runs from $(-\infty, \infty)$.  Let us denote the algebra of all observables localized in boundary time $t \in [0, \epsilon]$ by $\abdry$.   Then a statement of precisely the form above can be established. If $\rho_1$ and $\rho_2$  are any two states in the theory then imposing the equality of the expectation value of all elements in $\abdry$ imposes the equality of the states just as in \eqref{flatholinfo}. Such a statement would not be true in a nongravitational theory, where the algebra of boundary observables would be much larger than the algebra $\abdry$.

The result above is proved as ``Result 5'' in \cite{Laddha:2020kvp}. Note that this statement is {\em not} proved using AdS/CFT. It can be independently derived by a consideration of the operator algebra. In perturbation theory, this statement can also be derived directly through an analysis of the constraints imposed on states by the Wheeler-DeWitt equation \cite{Chowdhury:2021nxw}. Within perturbation theory, it is also possible to construct a simple ``physical protocol'' to extract information about the bulk from a small time band on the boundary \cite{Chowdhury:2020hse}.

Let us now consider the physical interpretation of  \eqref{flatholinfo}. In asymptotically flat space, $\scrip$ can be thought of as a limit of Cauchy slices on which one can specify data for massless excitations. Therefore, in asymptotically flat space, \eqref{flatholinfo} tells us that the information available on this entire Cauchy slice is available near its boundary, which is precisely the principle of holography of information. Equation \eqref{flatholinfo} can be interpreted similarly in AdS. The algebra of operators at the conformal boundary in the time band $[0, \epsilon]$ is similar to the algebra of operators at the time $t = {\epsilon \over 2}$ but in the radial range $r \in [\cot{ \epsilon \over 2}, \infty)$.  Therefore, in AdS,  \eqref{flatholinfo} again tells us that information available anywhere on a Cauchy slice is available near its boundary. 

Proofs of the principle of holography of information have only been given in these cases, although it seems very plausible that this should be a general property of gravitational theories. For related discussion, we refer the reader to \cite{Marolf:2006bk,Marolf:2008mf,Jacobson:2019gnm}.  However, it is not currently known how this principle applies when spatial slices are compact.

\subsubsection{Failure of the split property \label{subsecfailure}}
We now return to the implication for the split property. We emphasize that our discussion here is limited to bounded regions that are surrounded by their complement as shown in Figure \ref{figbounded}. The discussion below does {\em not} apply when the region and its complement both extend to asymptotic infinity.

In quantum gravity, it is impossible to define local gauge-invariant observables and so the first subtlety in considering the split property is to define what one means by the algebra, $\alset$, associated with a region.  One way to proceed is as follows. Even in quantum gravity, it is possible to find  {\em approximately localized} observables that probe the physics in the region $R$. A common way to obtain such observables is simply to fix gauge. Another, more elaborate procedure, is to ``dress'' such observables by means of a gravitational Wilson line \cite{Donnelly:2015hta}. However, as the latter picture makes clear, such operators are ultimately nonlocal. Nevertheless, starting with a set of approximately local operator in the region $R$, it is possible to define an algebra by taking all polynomials of these operators as shown  in \eqref{alsetgen}. One may similarly associate an algebra, $\overline{\alset}$ with the region $\rcomp$. We can then ask whether these algebras obey some version of the split property.

To answer this question, we note that whatever procedure is adopted for associating an algebra with a region it is clear that, in {\em all cases},  the algebra $\overline{\alset}$ must contain the algebra of all asymptotic operators that we denoted by $\abdry$ above. This simple observation immediately implies that the split property must fail in gravity.

This can be easily shown by contradiction. Assume that \eqref{splitproperty} holds in gravity. Then taking $A = 1$ and $\bar{A} \in \abdry$ we find that
\be
\tr(\rho \bar{A}) = \tr(\rho_2 \bar{A}), \quad \forall \bar{A} \in \abdry.
\ee
But then applying \eqref{flatholinfo} we see that we must have $\rho = \rho_2$. This means that for some nontrivial $A$ and for a general $\rho_1$ it is impossible to satisfy \eqref{splitproperty}.  Therefore, as claimed, it is impossible to specify the state of the theory inside and outside a bounded region independently in a theory of gravity.

 One important consequence of this result is that there is no simple definition of the entanglement entropy of the complement of a bounded region in gravity.  A natural question  is whether we can somehow coarse grain the set of observables  so as to recover an approximate split property and thereby define an entanglement entropy. We return to this question below although, in general, the answer is state dependent.

But, in many cases, the result \eqref{flatholinfo} is important even at leading order in perturbation theory. One striking example is as follows. 
Consider the vacuum of global AdS. Then,  as reviewed in \cite{Chowdhury:2020hse}, even if we keep track of the metric-degrees of freedom to leading nontrivial order in perturbation theory in ${l_{\text{pl}} \over l_{\text{AdS}}}$ (Planck scale divided by the AdS scale), the entanglement entropy of the complement of  a bounded region is trivial. This is because the measurement of the energy from near the boundary is enough to uniquely identify the global vacuum that is the only state with vanishing energy. The entanglement entropy of bounded regions and their complements is often studied in the vacuum of nongravitational field theories but such a quantity appears subtle to define even perturbatively in gravity in global AdS!

In other states, it is harder to extract information about a bounded region from its complement. For instance, it is nonperturbatively difficult to determine a black hole microstate by observations outside the black hole. This allows for a definition of a coarse-grained entropy for the exterior, even in the presence of gravity. Nevertheless, as we describe in section \ref{secpage},  the failure of the split property becomes relevant again if one attempts to define a fine-grained entropy that obeys a Page curve.

\section{The error in Hawking's argument \label{sechawkingerror}}
It is commonly stated that to resolve Hawking's original paradox, it is necessary to find the precise error in Hawking's original argument \cite{Hawking:1976ra}. We now explain how the failure of the split property in gravity allows us to isolate this error. 

In \cite{Hawking:1974sw} and \cite{Hawking:1976ra}, Hawking presented both a calculation and an argument. The calculation was as follows. Let $\rho$ be the final state of a black hole formed from collapse and let $a_{\omega}^{\dagger}, a_{\omega}$ be creation and annihilation operators that describe outgoing modes of frequency $\omega$ for a dynamical field.  Then \cite{Hawking:1974sw} showed that
\be
\label{hawkingrad}
\tr(\rho a_{\omega} a^{\dagger}_{\omega}) = {1 \over 1 - e^{-\beta \omega}},
\ee
where $\beta$ is the inverse temperature of the black hole. It was also argued that the occupancy of higher powers of $a_{\omega} a^{\dagger}_{\omega}$ would be thermal. (Here we have slightly smeared the operators to avoid writing a delta function in frequency as explained near equation 2.21 of \cite{Raju:2020smc}.)

By itself,  the result \eqref{hawkingrad} does {\em not} amount to a computation of the entanglement entropy  of Hawking radiation or an argument that the final state, $\rho$, must be thermal. This is because low-point correlators are insufficient to fix the entanglement entropy. For instance, if we were to modify the result above to
\be
\label{smallcorr}
\tr(\rho a_{\omega} a_{\omega}^{\dagger}) = {1 \over 1 - e^{-\beta \omega}} + \Or[e^{-{S_{\text{BH}} \over 2}}],
\ee
with corrections that are exponentially suppressed in the black hole entropy, $S_{\text{BH}}$,  this would be perfectly consistent with $\rho$ being a pure density matrix \cite{Lloyd:2013bza,Papadodimas:2012aq}.  Therefore  the result \eqref{hawkingrad} is not precise enough, by itself, to conclude that $\rho$ must be mixed.

But Hawking also provided a general argument to suggest that the purity of the initial state could not be preserved by small corrections of the form \eqref{smallcorr}. This argument was enunciated through, what was termed, the ``principle of ignorance'' in \cite{Hawking:1976ra}. In  \cite{Hawking:1976ra}, Hawking argued that it was necessary to introduce a ``hidden surface'' about a black hole background.  Hawking then stated that the  principle of ignorance would imply that (see p. 2463)
\begin{quote}
``all field configurations on these hidden surfaces are equally probably provided they are compatible with the conservation of mass, angular momentum, etc. which can be measured by surface integrals at a distance from the hole.''
\end{quote}

The argument of \cite{Hawking:1976ra} proceeded by explicitly introducing a factorization of the final Hilbert space. If $H_1$ is used to denote the Hilbert space describing data on the initial surface then it was argued that transition amplitudes would require the specification of a state on the ``hidden surface'' Hilbert space $H_2$ and a specification of the state on the final surface outside the black hole $H_3$. (See unnumbered equation on p. 2463 of \cite{Hawking:1976ra})  The crux of the argument in \cite{Hawking:1976ra} was that since data on the hidden surface would be lost to the outside observer, the outside observer would perceive a mixed state. 

\begin{figure}
\begin{center}
\includegraphics[width=0.4\textwidth]{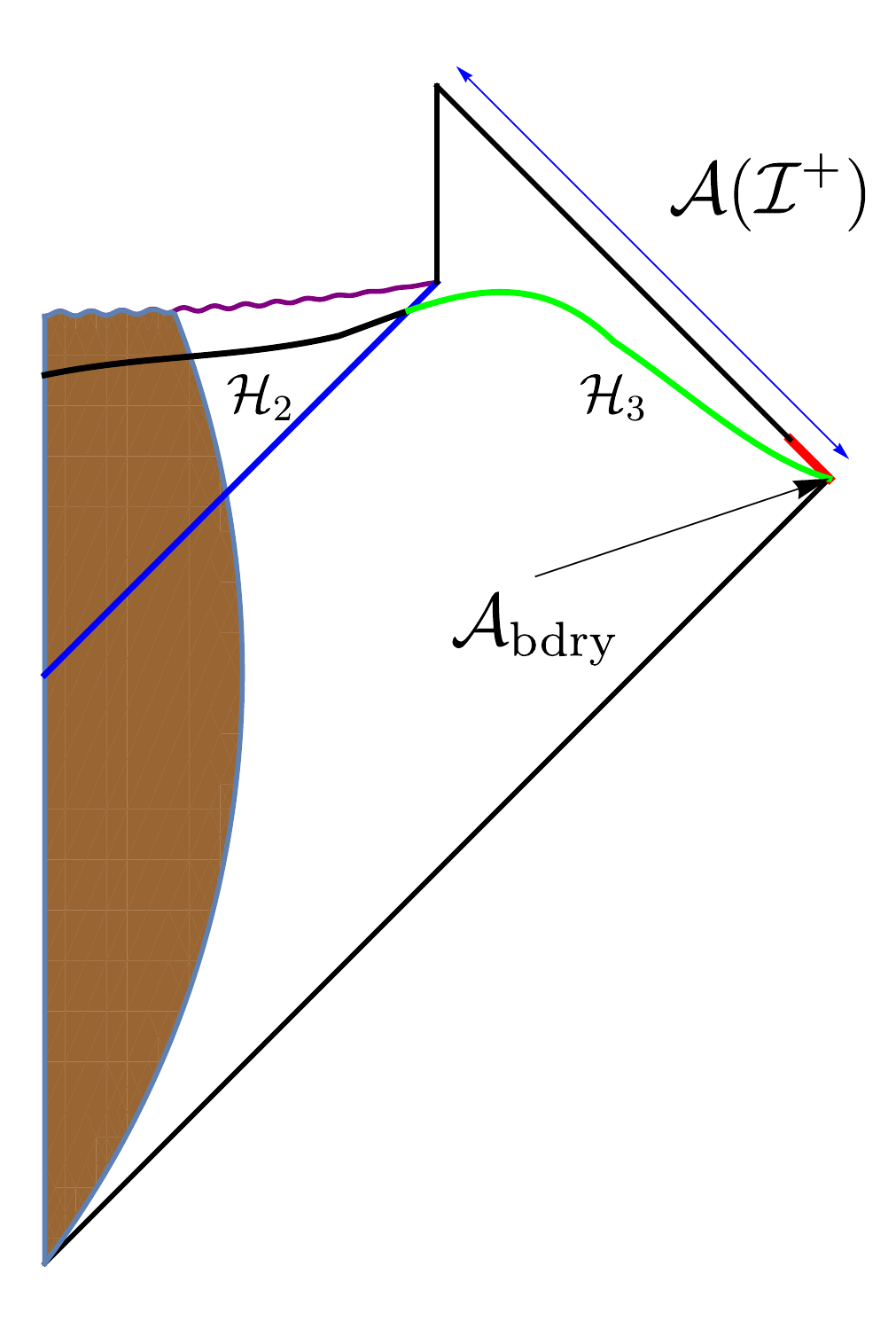}
\caption{\em The extended Penrose diagram of an evaporating black hole.  A common incorrect assumption is that the Hilbert space factorizes into a part $H_2$ associated with the interior and a part $H_3$ associated with the exterior, up to constraints imposed by mass and global charges.  But the Hilbert space does not factorize on a nice slice or even on ${\cal I}^{+}$. Information available in the algebra of all operators on ${\cal I}^{+}$ (marked in blue) is available in the algebra ${\cal \alset}_{\text{bdry}}$ of operators from the red region near ${\cal I}^{+}_{-}$. \label{figbhevap}}
\end{center}
\end{figure}
Figure \ref{figbhevap} restates this argument in the context of a nice slice through an evaporating black hole spacetime.  In a nongravitational quantum field theory, it is possible to specify data independently on the part of the nice slice behind the horizon and the part outside the horizon. Subject to the ultraviolet subtleties described in section \ref{secsplitnongrav}, the Hilbert space factorizes into a part $H_2$ associated with the interior and a part $H_3$ associated with the exterior. In classical gravity, it is possible to specify different configurations in the black hole exterior subject to only the constraints of some conserved charges. So if the split property were to hold in gravity, an observer at $\scrip$ would lose information about the state of the interior and would  have to trace over all possibilities for $H_2$ leading to a final mixed state at $\scrip$.

So we see that the argument of  \cite{Hawking:1976ra} relies explicitly on the split property  whose failure in theories of gravity was reviewed above. As explained in section \ref{secsplitfailure}, if one examines the full wavefunctional of the theory then it is {\em not} possible to specify data outside and inside the black hole separately. This quantum mechanical effect cannot be neglected if we ask fine grained questions about whether or not the final state is pure. 

To summarize, there are two specific issues with Hawking's argument for information loss.
\begin{enumerate}
\item
The computation of mode-occupancy performed by Hawking is not precise enough to conclude that the final state is mixed.
\item
In order to extend the validity of this calculation, Hawking invoked a ``principle of ignorance'' that explicitly assumed a factorization of the Hilbert space. The principle of holography of information implies that this assumption is incorrect in gravity.
\end{enumerate}

\section{The monogamy paradox  \label{secmonogamy}}
The original formulation of the information paradox was refined by Mathur in \cite{Mathur:2009hf} and Mathur's argument was later elaborated by   AMPS \cite{Almheiri:2012rt}. The idea of  \cite{Mathur:2009hf,Almheiri:2012rt} was to set up a potential contradiction between effective field theory, unitarity and the monogamy of entanglement that would be robust against small corrections of the form \eqref{smallcorr}.  A detailed review is provided in \cite{Raju:2020smc} and we provide a brief summary here to elucidate how the split property was assumed in these paradoxes.

Consider a nice slice in an evaporating black hole spacetime, as shown in Figure \ref{figbhevap} and divide it into three parts: $A,B,C$ where $A$ is just outside the horizon, $B$ is just inside the horizon and $C$ is a region near infinity corresponding to an asymptotic  observer.  We will repeat the original argument of \cite{Mathur:2009hf}, which assumed that, even in gravity, it should be possible to define an entanglement entropy corresponding to these regions and we denote the entropy of the region $R$ by $\sent_{R}$ below.

Effective field theory tells us that pairs of Hawking quanta in regions $A$ and $B$ must be entangled with each other, which led \cite{Mathur:2009hf} to the conclusion that
\be
\label{smoothorizon}
\sent_{AB} < \sent_{B}, \qquad \text{assuming~a~smooth~horizon},
\ee
i.e. the region B by itself is more ``mixed'' than $A$ and $B$ taken together. 

The argument in \cite{Mathur:2009hf} then proceeded by noting that if the entanglement entropy of the radiation follows a Page curve then for an old black hole the entropy of $C$ must start falling with time. But since the radiation in $C$ at a given time is the union of the radiation in $A$ and in $C$ at an earlier time we must have
\be
\label{oldbh}
\sent_{A C} < \sent_{C}, \qquad \text{assuming~a~Page~curve~for $\sent_C$~and~a~old~black~hole}.
\ee
These two claims are in contradiction with the strong subadditivity of entropy that reads
\be
\label{sse}
\sent_{AB} + \sent_{AC} \geq \sent_{B} + \sent_{C}.
\ee
In  \cite{Mathur:2009hf}, this was used this to argue that \eqref{smoothorizon} must break down and the interior of the black hole must be replaced by a ``fuzzball'' and in \cite{Almheiri:2012rt} the same argument was used to suggest the presence of ``firewall'' at the horizon.

But, a simple resolution to this paradox is obtained through the following observation. The strong subadditivity inequality \eqref{sse} holds {\em only} if the  operator algebras corresponding to subsystems $A,B,C$ commute.   Therefore, we see that a key assumption --- unstated but tacit --- that goes into the monogamy paradox of \cite{Mathur:2009hf} and \cite{Almheiri:2012rt} is that the algebra of operators can be separated into commuting factors associated with bounded geometric regions.  As explained in section \ref{secsplitfailure} this assumption is false in theories of gravity. This failure cannot be ignored  if we wish to make reference to fine-grained quantities like the entanglement entropy. So the purported refinement of Hawking's original paradox is based on the same erroneous assumption as the original paradox.

In the context of the discussion of the firewall, the issue of factorization was discussed in \cite{Bousso:2012as,Susskind:2012uw} and it was then observed in  \cite{Papadodimas:2012aq} that the paradox could be resolved by recognizing the failure of the independence of operator-algebras in gravity. Similar ideas were explored in \cite{Jacobson:2012gh}.

Although, as presented above,  the original paradox of  \cite{Mathur:2009hf} and \cite{Almheiri:2012rt} was phrased using the entanglement entropy,  it is possible to rephrase the paradox in terms of Bell correlators.  These correlators avoid the ambiguity in the definition of the entanglement entropy and can be defined in a theory of gravity. This makes the paradox sharper, although the resolution remains precisely the same as the one outlined above: the operator-algebra of the theory does not contain commuting factors associated with $A,B,C$. This reformulation and resolution can be seen particularly cleanly in a toy version of the paradox both in AdS \cite{Raju:2018zpn} and in flat space \cite{Chakraborty:2021rvy}.

\subsection{\bf Other paradoxes}
The original firewall paradox should be distinguished from subsequent paradoxes outlined in \cite{Almheiri:2013hfa,Marolf:2013dba}. The question addressed in the  paradoxes of \cite{Almheiri:2013hfa,Marolf:2013dba} is whether ``typical states in AdS/CFT have a smooth interior.'' These paradoxes are reviewed and sharpened in \cite{Papadodimas:2015jra}. They cannot be resolved through the simple observation that the Hilbert space does not factorize in quantum gravity. These paradoxes, and similar paradoxes that arise for the eternal black hole \cite{Papadodimas:2015xma} can be addressed through a state-dependent reconstruction of the black hole interior. The issue of the consistency of state dependence is further discussed in \cite{Marolf:2015dia,Raju:2016vsu}.  However, it is not clear if these issues are directly relevant for evaporating black holes since evaporating black holes are never typical states in the Hilbert space as discussed in \cite{Raju:2020smc}.

\section{The Page curve \label{secpage}}
We now turn to the common idea that the entanglement entropy of the region outside the black hole should follow a Page curve.  We explain why this idea also involves the same  assumption about the split property.

This assumption is stated explicitly in the original argument provided by Page and Lubkin \cite{Page:1993df,lubkin1978entropy}. The argument of \cite{Page:1993df} explicitly considered a factorized Hilbert space
\be
H = H_A \otimes H_B; \qquad \text{dim}(H_A) = n; \qquad \text{dim}(H_B) = m.
\ee
It was then argued that for  $|\Psi \rangle$  a pure state  in the large Hilbert space, the entropy of the reduced density matrix of a subsystem
\be
\rho_A = \tr_B |\Psi \rangle \langle \Psi |
\ee
would obey
\be
-\langle \tr_{A} (\rho_{A} \log \rho_{A}) \rangle_{\text{Haar}} = \text{min}(\log(n),\log(m)),
\ee
up to corrections suppressed by $\text{min}(n,m)/\text{max}(n,m)$ and where the expectation value is taken with respect to the Haar measure on pure states of the larger system. Since almost all states are Haar-typical this implies that for a ``generic pure state'' the entropy of a subsystem is given by the formula above.

The argument applies perfectly well to nongravitational systems and has been useful in that context \cite{nadal2010phase}. However, this argument is also often applied to an evaporating black hole. In \cite{Page:1993wv},  where this result was first applied to black holes (see paragraph above Eqn. 1 in the published version), Page assumed  that 
\begin{quote}
``the black hole subsystem  has dimension $n$ $\ldots$ the radiation subsystem has dimension $m$ $\ldots$ [and] these subsystems form a total system in a pure state in a Hilbert space of dimension $m n$.'' 
\end{quote}

If one could indeed treat the black hole and radiation as parts of a larger factorized Hilbert space,  the assumptions of \cite{Page:1993df,lubkin1978entropy} would be met, and the entanglement entropy of the radiation would follow a Page curve.

However, this is yet another restatement of the flawed assumption that we have discussed above. As explained in section \ref{secsplitfailure}, if one thinks of the black hole as occupying a bounded region on a Cauchy slice, and the radiation as occupying the complementary region then the state of the black hole is completely fixed once we specify all observables in the radiation region.  
Therefore, at least if we consider the fine-grained entropy, there is no reason to expect that the entropy of the radiation region  will follow a Page curve.

It is not even true that the Page curve appears at $\scrip$. Let $\alset(-\infty, u)$ denote the algebra of observables at $\scrip$ of the segment that extends from its past boundary to some cut at retarded time $u$. Then, as is standard \cite{Casini:2013rba}, one may define the entanglement entropy $\sent(u)$ with respect to this algebra. It can then be shown (see Result 3 in \cite{Laddha:2020kvp}) that
\be
{\partial \sent(u) \over \partial u} = 0,
\ee
which is the statement that the entanglement entropy is constant along $\scrip$ and does not obey a Page curve.

\subsection{The difference between ordinary statistical systems and black holes}
At this point the reader may wonder why the effect above can be neglected for ordinary statistical systems: although we live in a world where gravity is presumably quantized, we still expect that the entanglement entropy of ordinary systems will obey a Page curve.

This can be understood as follows. The difference between two typical microstates of an ordinary statistical system with thermodynamic entropy $S$ is suppressed  by $\Or[e^{-{S \over 2}}]$ \cite{lloyd1988black}. This follows simply from the fact that the number of relevant microstates is given by $e^{S}$.  Therefore to meaningfully discuss the entanglement entropy of a subsystem, we need to keep track of its observables to at least this accuracy. 
 On the other hand, gravitational effects are suppressed by a factor of $G E^{d-2}$ in $d$ dimensions,  where $E$ is the typical energy scale associated with excitations in the system. For ordinary systems, we can consider a limit where $G E^{d-2} \rightarrow 0$ while $S$ remains finite. 

However, such a limit does not exist for black holes. For a black hole, if we take the energy of a typical excitation to be $E=T$, where $T$ is the Hawking temperature then, for a Schwarzschild black hole in $d$ dimensions we find that
\be
G E^{d-2} = {\alpha  \over S_{\text{BH}}}, \qquad \qquad \alpha  =  {\pi^{3-d \over 2} \left({d - 3 \over 4}\right)^{d-2} \over   2 \Gamma({d-1 \over 2})}
\ee
where $S_{\text{BH}}$ is the entropy of the black hole. So there is no simple limit, where it is possible to keep the entropy of the black hole finite while neglecting the gravitational effects described in section \ref{secsplitfailure}.

Even in a theory of gravity, there is no obstruction to thinking of the black hole radiation as a subsystem, in a coarse-grained sense. We can simply focus on observables that are insensitive to effects that are suppressed by powers of $S_{\text{BH}}$. For instance,  the classical expectation value of the metric and other geometric quantities are all of this form. However, within this approximation, it is not possible to speak meaningfully of a fine-grained entropy for the radiation. It is possible that there is some intermediate level of ``coarse graining'' where a Page curve appears for black holes, and we discuss this possibility in section \ref{subsecpageposs}.

\subsection{Computations of the Page curve}

We now turn to the computations of the Page curve in the recent literature \cite{Almheiri:2019psf,Penington:2019npb,Almheiri:2019hni,Almheiri:2019psy,Almheiri:2019yqk,Almheiri:2020cfm}. We first discuss the nature of ``information transfer'' in these computations. We then describe why these computations involve massive gravity, and their lessons are inapplicable to standard theories of gravity.

\subsubsection{Nature of information transfer}
In popular descriptions\cite{Musser:2020}, it is sometimes stated that the recent computations of the Page curve indicate how information ``emerges'' from a black hole. However,  this is misleading.

 Computations of the Page curve have been performed by coupling a holographic gravitational system in AdS to a nongravitational bath. Therefore the Page curve that is computed only measures how information is transferred from one section of this nongravitational system to another. These computations are perfectly consistent with the results explained above, which state that information is always available near the boundary of the gravitational spacetime.

\begin{figure}[!ht]
\centering
\begin{subfigure}{0.4\textwidth}
\centering
\includegraphics[height=0.3\textheight]{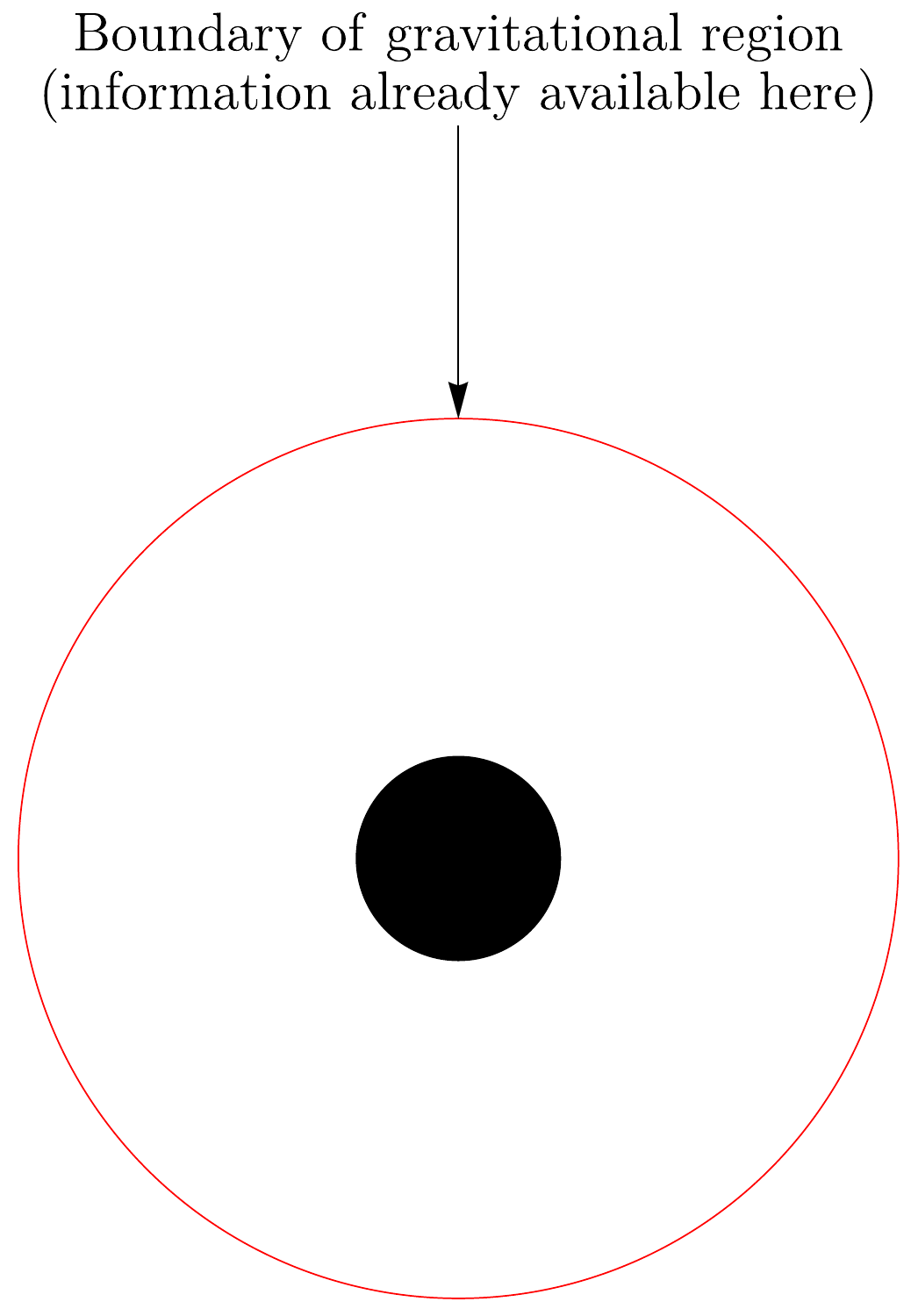}
\caption{\label{figtwostepa}}
\end{subfigure}
\hspace{0.15\textwidth}
\begin{subfigure}{0.4\textwidth}
\centering
\includegraphics[height=0.3\textheight]{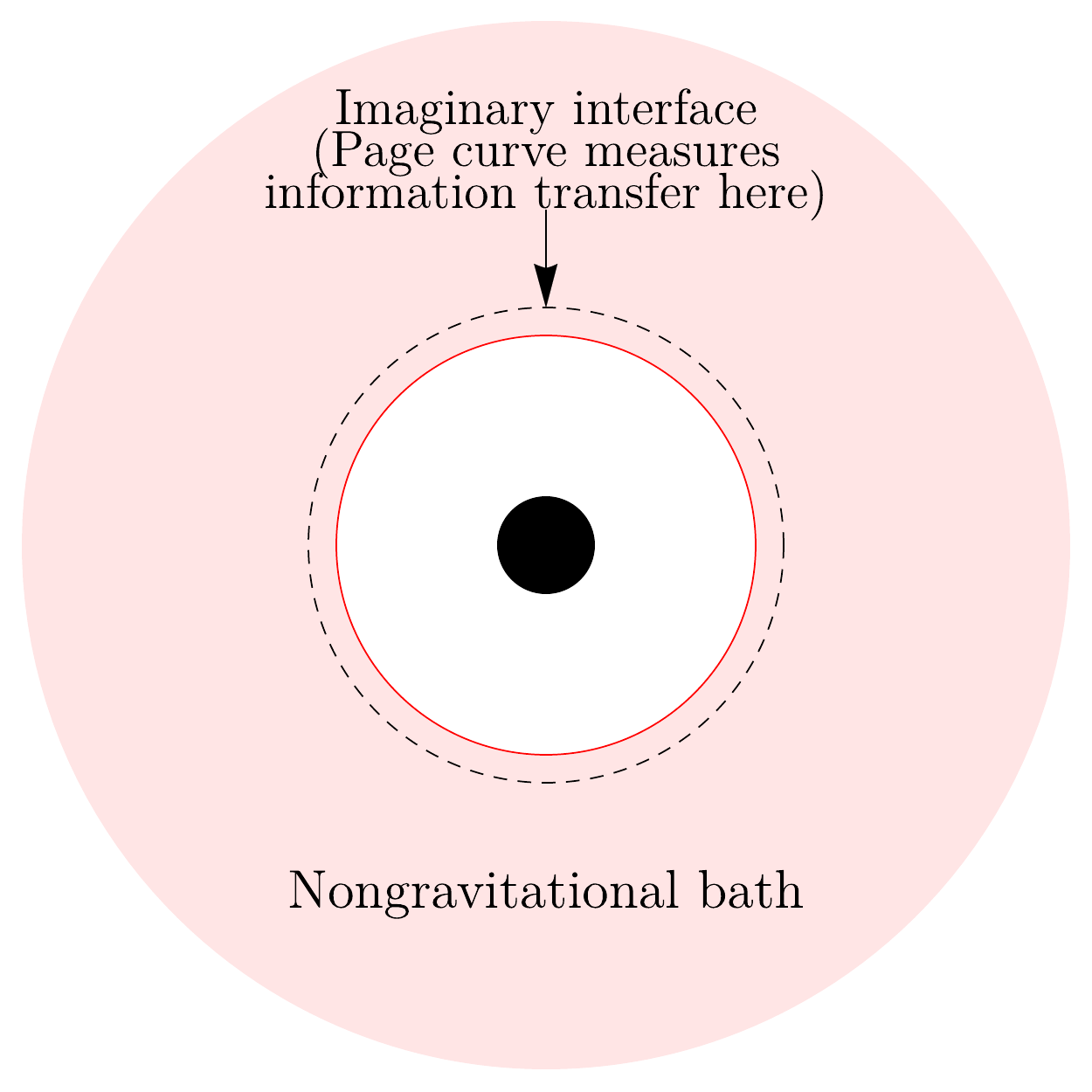}
\caption{\label{figtwostepb}}
\end{subfigure}
\caption{\em A two step process clarifying the nature of information transfer in computations of the Page curve. First, we prepare a black hole in AdS (left subfigure). According to the principle of holography of information, information about the microstate resides near the boundary of the spacetime. Second, we couple the system to a nongravitational bath (pink shaded region in the right subfigure). Information flows across a nongravitational interface (dashed line) and this information transfer is described by the Page curve. \label{figtwostep}}
\end{figure}

This can be understood more clearly by considering the two-step process shown in Figure \ref{figtwostep}. In the first step (Figure \ref{figtwostepa}) we prepare a black hole in a theory of a gravity in an asymptotically AdS spacetime. According to the discussion of section \ref{secsplitfailure}, all information about the state is always present near the boundary of the spacetime. In the second step (Figure \ref{figtwostepb}), we couple a nongravitational bath to the conformal boundary of AdS. Information then flows into the nongravitational bath. 

Since the bath is nongravitational, it does obey the ``split property''. Therefore, if one draws an imaginary interface in this bath, the information transfer across this interface is controlled by a Page curve as one would expect on general grounds in any nongravitational system. This is the Page curve that has been computed.

Moreover, the computation of this Page curve is also unnecessary to establish the unitarity of the original process of black-hole formation.  If the black hole is formed from a pure state, unitarity follows from the fact that the entropy of the boundary remains zero after the black hole forms. The Page curve is a check that the theory remains unitary after it is coupled to a bath. But this latter question is a question about the nongravitational bath and its coupling to the original system. This question is interesting in its own right but it is not a question that is important for the original gravitational theory.

Several computations of the Page curve use  ``doubly holographic'' settings where the system of the black hole and bath can be embedded in a higher-dimensional AdS. This is achieved by inserting a Karch-Randall \cite{Karch:2000gx,Karch:2000ct} brane in AdS, and the entropy of a part of the bath can then be studied by the standard Ryu-Takayanagi prescription \cite{Ryu:2006ef,Ryu:2006bv,Hubeny:2007xt}. (See \cite{Almheiri:2019psy,Chen:2020uac,Uhlemann:2021nhu} for an elaboration of this perspective.) The authors of \cite{Geng:2020fxl} generalized this setup by considering an AdS space with two Karch-Randall branes inserted in a black-string geometry. This serves as a model of a system where the bath itself is gravitating. 

It was then found in \cite{Geng:2020fxl} that there was no nontrivial Ryu-Takayanagi surface (excluding the black-string horizon itself) that extended from one brane to the other with symmetric Neumann boundary conditions imposed on both branes. Therefore, the Page curve disappears when the bath itself is gravitating.  In \cite{Geng:2020fxl}, it was shown that this conclusion is true even if gravity on the ``bath brane'' is much weaker than gravity on the ``system brane''. This lends evidence to the idea that systems with ``weak gravity'' should not be conflated with systems with ``no gravity.''

\subsection{Massive gravity and islands.}
As explained above, extant Page curve computations address a nongravitational problem. What is interesting, nevertheless, is that this nongravitational problem can be studied using gravitational techniques, including an elegant ``island rule'' \cite{Almheiri:2020cfm}, that give answers consistent with the Page curve.

While the emergence of the Page curve in this nongravitational context is natural, a closer examination of the gravitational computation reveals that the ``island rule'' appears to be in tension with the principle of holography of information. The island rule suggests that after the bath has been coupled to the holographic system, the state of an ``island'' inside the gravitational system, is captured by part of the bath and the state of its complement in the gravitational region is captured by the complementary part of the bath.  Since the nongravitational bath obeys a split property, this implies that it is also possible to set up a split state in the gravitational bulk. This seems to be in contradiction with the result of section \ref{secsplitfailure}.

The authors of \cite{Geng:2021hlu} turned this into a sharp puzzle for theories of gravity in higher than $1+1$ dimensions. This puzzle  does not require the full power of the principle of holography of information. Rather, it can be shown, using only  perturbation theory, that islands are inconsistent in a theory of gravity that obeys the Gauss law. 

The puzzle established in \cite{Geng:2021hlu} exploits the observation that correlators of simple operators in a typical black hole microstate, $|\Psi_{\text{typ}} \rangle$ are approximately time independent. Now consider a black hole that contains an island, and consider creating an excitation in the island using an operator of the form $U = e^{i \lambda \phi(P)}$, where $\phi(P)$ is a weakly coupled propagating field smeared near a point $P$ in the island. A simple perturbative computation reveals that
\be
\label{changelambda}
{\partial \over \partial \lambda}  \langle \Psi_{\text{typ}} | U^{\dagger} H {\partial \phi(P') \over \partial t} U | \Psi_{\text{typ}} \rangle \neq 0,
\ee
where $H$ is the integral of the metric near the AdS boundary that measures the total energy and $\phi(P')$ is an insertion of the same scalar field smeared near a point $P'$ at the asymptotic boundary and $t$ is the boundary time. The right hand side can be computed explicitly in perturbation theory. (See section  3 in \cite{Geng:2021hlu}.)  But this contradicts  the idea that simple unitary operators in an entanglement wedge should leave correlators outside the wedge invariant. 

Note that in equation \eqref{changelambda} one is {\em not} attempting to identify the original black hole microstate, which would not be possible within perturbation theory. Rather, equation \eqref{changelambda} exploits the fact that an excitation on top of a typical microstate leads to an atypical state. It is  the difference between this atypical state and the original typical state that can be detected in perturbation theory by simple correlators outside the horizon \cite{Papadodimas:2017qit}.

The physical observation underlying the puzzle is simply that islands are unique among entanglement wedges because islands are surrounded by their complements. This brings islands within the purview of section \ref{secsplitfailure}; in contrast, standard entanglement wedges in AdS/CFT always contain a piece of the asymptotic region and so the discussion of section \ref{secsplitfailure} does not apply to them.  In standard entanglement wedges, it makes sense to consider a little Hilbert space (as introduced in \cite{Papadodimas:2013jku} or, equivalently, a ``code subspace'' \cite{Almheiri:2014lwa}) and one can study excitations within the entanglement wedge that leave all correlators in the complement of the wedge invariant. This is because the excitation can be ``dressed'' to infinity through the asymptotic part of the wedge. Since islands do not contain such a region, in a standard theory of gravity, there are no excitations that leave the complement of the island invariant. So, in standard gravity, islands cannot constitute entanglement wedges, at least in the usual sense of the term.  The computation \eqref{changelambda} shows this explicitly.

This does not mean that computations based on the island rule are wrong. Rather the resolution to the puzzle is that (in higher than $1+1$ dimensions)  the coupling the gravitational system to a bath causes the bulk graviton to pick up a mass. Therefore almost all computations based on islands have been carried out in massive gravity \cite{Geng:2020qvw}. Massive gravity is quite different from standard gravity; it does not have a Gauss law, and so the principle of holography of information does not apply and islands are consistent in such theories. Before we explain how this resolves the puzzle we first explain why the nongravitational bath induces a mass for the graviton.

The easiest way to understand the graviton mass \cite{Aharony:2006hz} is to again consider the sequence shown in Figure \ref{figtwostep}, where a black hole is first prepared in a holographic system and later coupled to a bath. In Figure \ref{figtwostepa}, before the system is coupled to a bath, the holographic stress tensor on the boundary of AdS is conserved: 
\be
\partial_{\mu} T^{\mu \nu} = 0, \qquad \text{(in~Figure~\ref{figtwostepa}~before~coupling~to~bath)}.
\ee
where $\mu, \nu$ run along the boundary coordinates. After the coupling is turned on, we now have 
\be
\partial_{\mu} T^{\mu \nu} \neq 0.  \qquad \text{(in~Figure~\ref{figtwostepb}~after~coupling~to~bath)}.
\ee
This must be true since, by construction,  the coupling allows energy to leak into the bath.   

In conformal representation theory \cite{Mack:1975je,Minwalla:1997ka,Dobrev:2004tk,Dobrev:2002dt}, the conservation of the stress tensor is simply the statement that the stress tensor lives in a ``short representation'' of the conformal algebra. This allows its conformal dimension to be exactly $d$ where $d$ is the boundary dimension. The non-conservation of the stress tensor implies that, in the theory with the bath, the stress tensor is in a ``long representation'' but this means that its conformal dimension must exceed $d$, and so it must pick up an anomalous dimension. By the standard rules of AdS/CFT \cite{Witten:1998qj,Gubser:1998bc} this means that the bulk graviton must have a mass. This can be verified through a direct bulk computation \cite{Porrati:2001gx,Porrati:2003sa}.

The constraints of massive gravity are different from those of standard massless gravity. Most importantly, the Hamiltonian of the theory of massive gravity is {\em not} a boundary term. Therefore, the principle of holography of information does not apply and even the computation leading up to \eqref{changelambda} is modified. 

The simplest setting to see this difference is to compare linearized standard gravity with linearized massive gravity. Writing the metric on a spatial slice as $\delta_{i j} + h_{i j}$, we find that the metric fluctuation is constrained in terms of the local energy density, $\rho$ as
\be
{1 \over 16 \pi G} \left( \partial_j \partial_i h_{i j} -\partial_j \partial_j h_{i i} \right)  = \rho,
\ee
in standard gravity. Upon integration this leads to the familiar ADM expression \cite{Arnowitt:1962hi} for the total energy as a surface integral.
\be
\label{admhamilt}
H = {1 \over 16 \pi G} \int_{S_{\infty}} \ n_j(\partial_i h_{i j} - \partial_j h_{i i} ),
\ee
where the integral is performed over the sphere at asymptotically large $r$ with normal $n_j$. Of course,  \eqref{admhamilt} is the {\em definition} of the energy even when one goes beyond the linearized approximation to the full theory of general relativity (see section 20.4 in \cite{misner1973gravitation}) and also in the quantum theory \cite{Regge:1974zd}.  On the other hand, in linearized massive gravity we find that the analogous constraint is \cite{Hinterbichler:2011tt} 
\be
{1 \over 16 \pi G} \left(\partial_j \partial_i h_{i j}-\partial_j \partial_j h_{i i}  + m^2 h_{i i} \right) = \rho.
\ee
Since the left hand side is not a total divergence, the integral of the energy density is not captured by an integral at infinity even in the linearized theory.

In \cite{Geng:2021hlu}, it is shown that, at least, in some well-understood examples of massive theories of gravity, such as those that appear in the Karch-Randall model, there does not appear to be any any immediate obstacle to preparing split states.\footnote{Here, we should caution the reader that the question of how information is localized in theories of massive gravity has not been studied in detail.}

To summarize: existing derivations of the Page curve pertain to information transfer across a nongravitational interface and do not indicate the transfer of information within the gravitational region.  Moreover, the models that have been used to compute the Page curve involve rather nonstandard theories of gravity. These models differ qualitatively from standard gravity, and it does not appear that the island proposal or other results from such models can be generalized to black holes in standard gravity.

\subsection{Page curves in standard gravity \label{subsecpageposs}}
In the section above, we criticized the expectation that the radiation from a black hole in standard gravity should follow  a Page curve.
However, since this question has historically  been of interest, one may ask whether it is possible to find an appropriate quantity,  in a standard theory of gravity, that does follow a Page curve. Here, we outline some possibilities in this direction.

\paragraph{Page curves in flat space.}
First, consider four-dimensional flat space. Then the algebra of operators at $\scrip$ that we denoted in section \ref{secsplitfailure} as $\alset(\scrip)$  can be separated into an algebra of ``dynamical operators'' (generated by the so-called ``Bondi news'' operators and tails of other massless degrees of freedom at $\scrip$) and a set of ``constrained operators'', which include the ADM Hamiltonian and other soft charges. These constrained modes are what cause the split property on $\scrip$ to fail. On the other hand, since the operator algebra of dynamical modes at $\scrip$ is otherwise free, arbitrary products of dynamical operators do not generate constrained operators. The algebra of dynamical operators can therefore be separated into an algebra corresponding to a segment of $\scrip$ and a commuting algebra corresponding to its complement.

This is still not sufficient to ensure that the entropy with respect to the dynamical algebra follows a Page curve; it is necessary to further assume that, in a generic state,  the entanglement between the hard modes and the different soft vacua \cite{Strominger:2013jfa,Ashtekar:2018lor}  can be neglected. The validity of this assumption has not been investigated in detail and the proposal of \cite{Hawking:2016msc} was that this entanglement is significant. But if this assumption holds then it is expected that the entropy of a state with respect to the coarse-grained algebra above will follow a Page curve.

Although the coarse-graining above is mathematically consistent, it is somewhat artificial from a physical point of view. The observer must be sensitive to very small fluctuations of the metric, including those with a magnitude $\Or[e^{-S}]$ but is not allowed to include the mass of the black hole and its correlators with other degrees of freedom in the algebra of observables. 

There are other possibilities for obtaining a Page curve at $\scrip$ all of which boil down to the coarse-graining above. These include considering the algebra of operators in the retarded-time interval $(u, \infty)$ --- which extends from a cut at finite retarded time till the future boundary of future null infinity --- or equivalently considering the algebra of operators in the infinitesimal retarded-time interval $(u, u + \epsilon)$. We refer the reader to section 5.4.5 of \cite{Raju:2020smc} for additional discussion.

\paragraph{Page curves for small AdS black holes.}
It is also interesting to consider small black holes in AdS \cite{Lowe:1999pk}. If the radius of the black hole is much smaller than the AdS scale, it will evaporate entirely with no need to couple the system to an external bath. Of course, in this case we know from AdS/CFT --- and entirely consistent with the discussion in section \ref{secsplitfailure} --- that information about the black-hole microstate is present at the boundary of AdS at all times. Therefore if the black hole is formed from the collapse of a pure state, the fine-grained entropy of the boundary remains zero at all times.

On the other hand, if one considers a small set of coarse-grained observables --- such as generalized free-fields and products of a small number of such operators --- then one expects to see an entropy that simply rises according to Hawking's original prediction. 

In order to see a Page curve, one needs to find an intermediate level of accuracy --- an algebra that keeps enough information that it can detect that the state is pure after the black hole has evaporated but not so much information that it knows that the state is pure at all times.  

Although, currently, there is no known subalgebra of the full boundary algebra that satisfies these properties, it appears plausible that such an ``intermediate-level'' algebra can be found. However, it is unclear whether such an algebra will be natural from a physical viewpoint or of intrinsic interest separate from reproducing the Page curve.

\section{Conclusion}

In this essay, we have pointed out how a familiar physical assumption is, in fact, false in quantum gravity. In nongravitational  theories, the algebra of observables can be separated into a subalgebra associated with a bounded region and another commuting subalgebra associated with its complement. This property fails in a theory of quantum gravity. 

For the purposes of defining a fine-grained entanglement entropy, this issue can be neglected for those systems where it is possible to take a limit in which the degeneracy of microstates remains finite while gravitational effects become arbitrarily weak.  However, the limit above does not exist for black holes, and so this gravitational effect is important if we study the fine-grained entanglement entropy of the radiation.  In a fine-grained sense, this entropy is trivial and the region outside a black hole always has information about the interior. 

This physical effect was not accounted for in Hawking's original formulation of the information paradox, which assumed that the Hilbert space in gravity should factorize the way it does in nongravitational theories. Our current understanding allows us to pinpoint the precise error in Hawking's argument.  Ironically, the same  assumption of factorization also underlies the expectation that the entropy of the radiation  should obey a Page curve.   

The point of this essay is not that it is impossible to find a Page curve for an evaporating black hole. Indeed, if we are willing to suitably change the question that is being addressed, it is possible to find an appropriate question to which the answer is the Page curve. In several extant computations in the literature, the Page curve has been derived by considering a nongravitational bath and then computing the entanglement between two regions of this bath. The gravitational description of this system involves a theory of massive gravity that appears to localize information like ordinary quantum field theories.  Even in standard theories of gravity, it appears plausible that a prescription for coarse-graining the entropy or somehow neglecting gravitational effects will cause a Page curve to emerge although such a computation has not been carried out.

However, the reason for studying the black-hole information paradox is that it teaches us about new physical effects in gravity. As such, one of the lessons that the paradox teaches us is that gravity localizes information unusually. This is a striking effect that persists in the low-energy theory. 

A computation of the Page curve is not necessary to resolve the information paradox and, historically, this idea was based on a flawed physical analysis.
Although it is possible to modify the system of an evaporating black hole in a standard theory of gravity so as to force a Page curve upon it,   this tends to obscure the interesting physics that we learn from the paradox.

\section*{Acknowledgments}
I am grateful to   Subhro Bhattacharjee, Tuneer Chakraborty, Joydeep Chakravarty, Chandramouli Chowdhury,  Hao Geng,  Victor Godet,  Andreas Karch,  Alok Laddha,  Carlos Perez-Pardavila,  Kyriakos Papadodimas, Olga Papadoulaki, Priyadarshi Paul, Siddharth Prabhu, Lisa Randall, Marcos Riojas,  Sanjit Shashi and Pushkal Shrivastava for helpful discussions.  This work was partially supported by a Swarnajayanti fellowship,  DST/SJF/PSA-02/2016-17, of the Department of Science and Technology. Research at ICTS-TIFR is supported by the government of India through the Department of Atomic Energy grant RTI4001.

\bibliographystyle{utphys}
\bibliography{references}
\end{document}